\documentclass[article]{IEEEtran}

\usepackage{graphicx}
\usepackage{amsmath}
\usepackage{amsfonts}
\usepackage{array}
\usepackage{booktabs}
\usepackage{adjustbox} 
\usepackage{multirow} 
\usepackage{xcolor, colortbl}
\usepackage{hhline}
\usepackage{makecell}
\usepackage{xcolor}
\usepackage{url} 
\usepackage[utf8]{inputenc} 

\ifCLASSOPTIONcompsoc
\usepackage[caption=false,font=footnotesize,labelfont=sf,textfont=sf]{subfig}
\else
 \usepackage[caption=false,font=footnotesize]{subfig}
\fi

\DeclareMathOperator*{\argmax}{arg\,max} 
\begin{document}

\title{Detecting Airborne Objects with 5G\,NR Radars}

\author{Steve Blandino$^{1,2}$ \thanks{This work is a U.S. Government work and not subject to U.S. copyright.}, Nada Golmie$^{3}$, Anirudha Sahoo$^{3}$, Thao Nguyen$^{3}$, \\Tanguy Ropitault$^{1,2}$,  David Griffith$^{3}$ and Amala Sonny$^{1}$ \\
$^{1}$ Associate, National Institute of Standards and Technology (NIST), Gaithersburg, MD \\
     $^{2}$  Prometheus Computing LLC, Bethesda, MD.\\
     $^{3}$  National Institute of Standards and Technology (NIST), Gaithersburg, MD.
     }


\maketitle

\begin{abstract}
The integration of sensing capabilities into 5G New Radio (5G\,NR) networks offers an opportunity to enable the detection of airborne objects without the need for dedicated radars.
This paper investigates the feasibility of using standardized Positioning Reference Signals (PRS) to detect UAVs in Urban Micro (UMi) and Urban Macro (UMa) propagation environments. A full 5G\,NR radar processing chain is implemented, including clutter suppression, angle and range estimation, and 3D position reconstruction. Simulation results show that performance strongly depends on the propagation environment. 5G\,NR radars exhibits the highest missed detection rate, up to 16\,\%, in UMi, due to severe clutter.
Positioning error increases with target distance, resulting in larger errors in UMa scenarios and at higher UAV altitudes. In particular, the system achieves a position error within 4\,m in the UMi environment and within 8\,m in UMa. The simulation platform has been released as open-source software to support reproducible research in integrated sensing and communication (ISAC) systems.
\end{abstract}


\IEEEpeerreviewmaketitle

\section{Introduction}

The integration of sensing capabilities into Fifth-generation New Radio (5G\,NR) and future networks have gained significant attention in recent years, positioning cellular infrastructure as a promising foundation for a wide range of sensing applications \cite{9829746}. Although 5G\,NR was originally designed to support high-throughput, low-latency communications, its waveform structure, beamforming capabilities, and high-resolution reference signals make it suitable for sensing tasks \cite{9921271,icc25}.

Sensing with cellular networks enables a variety of applications, including industrial automation, traffic monitoring, and environmental sensing. Among these, one of the most relevant and high-impact use cases, in the commercial and defense domains, is the detection and tracking of airborne objects such as unmanned aerial vehicles (UAVs) \cite{10938573}. The widespread use of UAVs in civilian and military applications introduces new challenges for airspace monitoring and security. Reliable detection and trajectory estimation of such sensing targets (STs) is critical for early threat identification and the prevention of unauthorized intrusions \cite{8337900}.

While traditional radar systems have proven effective for air defense, their deployment requires dedicated hardware installations and dedicated, licensed spectrum allocations, resulting in significant cost and complexity to achieve broad or nationwide coverage. In contrast, 5G\,NR infrastructure, already extensively deployed across both urban and rural areas, offers a cost-effective and scalable alternative. This opens the door to distributed sensing architectures, where 5G\,NR base stations (BSs) can complement dedicated military radar systems by providing real-time aerial threat detection and situational awareness. With such a system design, sensitive and critical areas would remain protected by dedicated radars, while broader regions would be continuously monitored through 5G\,NR-based sensing.

While 5G\,NR offers promising capabilities for airspace surveillance, its technical feasibility for UAV detection is yet to be fully investigated, particularly in terms of the maximum detection range and the impact of deployment conditions on sensing performance. Some recent studies have demonstrated the use of 5G\,NR signals for UAV detection, either through experimental validation~\cite{9695141,10172437} or electromagnetic simulation~\cite{10625724}. Others have proposed radar networks using 5G BSs to enable multistatic configurations for UAV detection~\cite{10615952}. However, the fundamental limits of detection and positioning performance, particularly in the presence of clutter and across diverse propagation scenarios, remain poorly understood. This paper addresses this gap by evaluating the ability of 5G\,NR radar systems to detect and localize UAVs under realistic environmental conditions, using 3GPP channel models and considering situations where line-of-sight (LoS) conditions are harder to achieve due to clutter, such as in urban and suburban deployments.

In this work, we focus on a monostatic sensing configuration, where a single 5G\,NR BS operates as both transmitter and receiver to detect and localize STs in its surrounding environment. The BS transmits 5G\,NR communication signals and receives echoes scattered by objects in the scene. The BS estimates the position of STs by processing the angle of arrival (AoA), and the time delay of the received signals.

To obtain sensing observations, we employ the positioning reference signal (PRS) defined in the 5G\,NR standard~\cite{3gppTS38211}, following the approach proposed in \cite{9921271}. We evaluate the PRS-based system’s detection performance in the presence of clutter and multipath components, which can impair reliable ST identification. Our analysis considers realistic operational scenarios, including urban micro (UMi) and urban macro (UMa). We implement path loss, shadow fading, and LoS probability models based on 3GPP technical reports TR~38.901~\cite{3gppTR38901} and TR~36.777~\cite{3gppTR36777}. Furthermore, taking into account ongoing 3GPP developments toward Release~19 for integrated sensing support, we incorporate recent ST radar cross-section (RCS) and monostatic background clutter models to reflect realistic sensing conditions.

To promote reproducibility and foster future research on joint communication and sensing with 5G\,NR infrastructure, we have publicly released our simulation platform as open-source software \cite{blandino2025}. 

The remainder of this paper is structured as follows: Section II describes the system model. Section III describes the 5G\,NR radar processing to achieve position estimates. Section IV presents the results, including positioning error analysis. Finally, Section V concludes the paper.
\section{System Model}
In this section, we present the system architecture and signal model adopted to evaluate the sensing capabilities of a 5G\,NR radar for ST detection. We consider a monostatic radar configuration where the BS simultaneously transmits and receives PRS, exploiting echoes to estimate ST parameters.

\subsection{5G NR Radar Model}

\begin{figure}
    \centering
    \includegraphics[width=1\columnwidth]{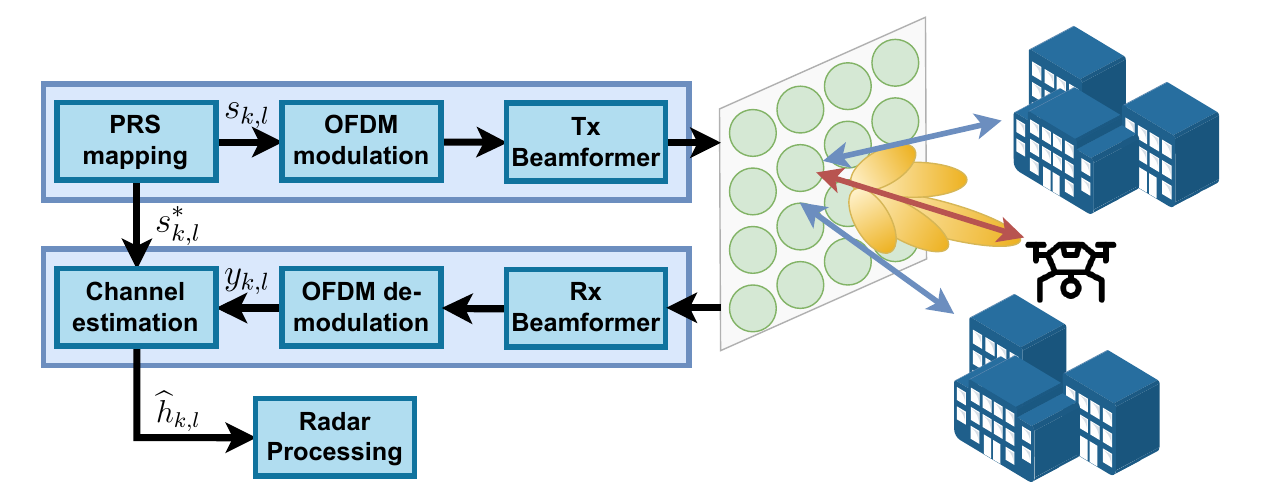}
    \caption{System model illustrating the monostatic 5G NR radar configuration with PRS-based sensing.}
    \label{fig:system_model}
\end{figure}
We consider a single 5G NR base station operating as an OFDM radar~\cite{braun2014ofdm}, as illustrated in Fig.~\ref{fig:system_model}. The BS is equipped with \( N \) co-located transmit and receive antennas, arranged in a uniform rectangular array enabling analog beamforming in both azimuth and elevation.

\begin{figure}
    \centering
    \includegraphics[width=1\columnwidth]{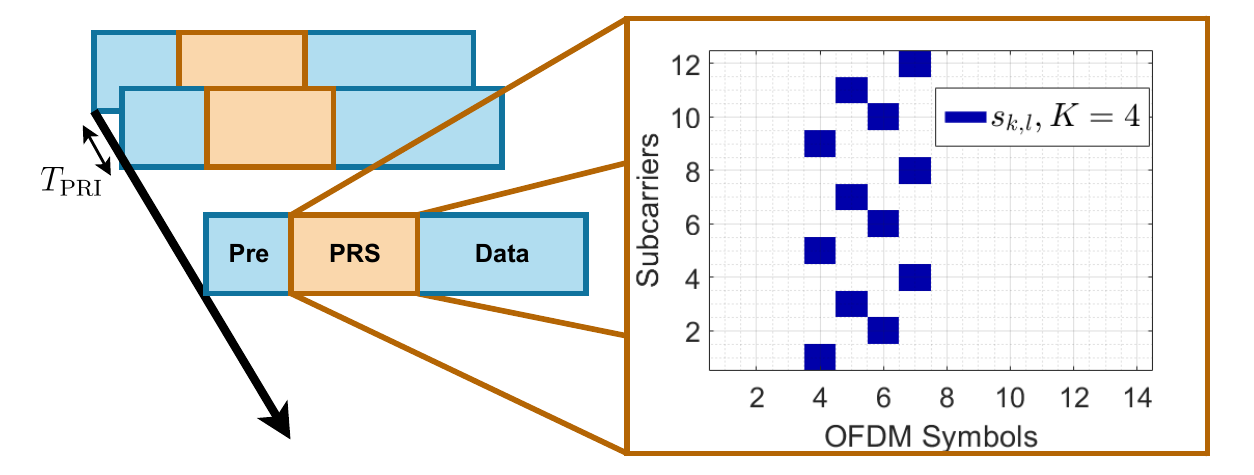}
    \caption{Example PRS resource grid structure, highlighting comb subcarrier mapping and PRS symbol allocation.}
    \label{fig:prs}
\end{figure}

For sensing, the BS transmits PRS sequences generated by Quadrature Phase-Shift Keying (QPSK)-modulating pseudo-random binary sequences, following the 31st-order Gold sequence specification described in  3GPP TS 38.211, Section 7.4.1.7.1 \cite{3gppTS38211}. The \( l \)-th PRS-OFDM symbol at subcarrier \( k \) is defined as:
\begin{equation}
s_{k,l} =
\begin{cases}
\beta_\mathrm{PRS} \, c_{k,l}, & \text{if } k \bmod K = 0 \\
0, & \text{otherwise},
\end{cases}
\end{equation}
where \( c_{k,l} \) denotes the QPSK-modulated PRS symbol, $K$ is the comb spacing, and \( \beta_\mathrm{PRS} \) is an amplitude scaling factor used to normalize the total transmit power. Since only every \( K \)-th subcarrier is active, the energy per active tone increases, necessitating power adjustment via \( \beta_\mathrm{PRS} \) to maintain constant symbol transmit power.
The subcarrier spacing is \( \Delta f \), leading to active tones spaced by \( K \Delta f \) across the bandwidth. In each slot configured for PRS transmission (termed a PRS occasion), \( L_\mathrm{PRS} \) OFDM symbols are allocated, following a comb-type resource mapping, as illustrated in Fig.~\ref{fig:prs}. According to 3GPP TS 38.211~\cite{3gppTS38211}, PRS supports four comb configurations with \( K \in \{2,4,6,12\} \), and each PRS occasion includes \( L_\mathrm{PRS} \in \{2, 4, 6, 12\}\)  PRS OFDM symbols, such that $L_\mathrm{PRS} \ge K$.
\( N_\mathrm{PRS} \) PRS occasions are scheduled periodically with repetition interval \( T_\mathrm{PRS} \), functionally equivalent to the Pulse Repetition Interval (PRI) in classical radar systems. 

The PRS grid is OFDM-modulated by applying an inverse discrete Fourier transform (IDFT) per OFDM symbol, followed by cyclic prefix insertion. After OFDM modulation, analog beamforming is applied. The baseband transmit signal at subcarrier \( k \) and OFDM symbol \( l \) can be expressed as:
\begin{equation}
\mathbf{x}_{k,l} = \mathbf{f}\, s_{k,l},
\end{equation}
where \( \mathbf{x}_{k,l} \in \mathbb{C}^{N \times 1} \) is the precoded vector and \( \mathbf{f} \in \mathbb{C}^{N \times 1} \) is the frequency-flat analog beamforming vector steering towards the ST.
At the receiver, leveraging the cyclic prefix which ensures circular convolution, the received symbol at subcarrier \( k \) and OFDM symbol \( l \) is:
\begin{equation}
y_{k,l} = \mathbf{w}^H \mathbf{H}_{k,l} \mathbf{f}\, s_{k,l} + n,
\end{equation}
where \( \mathbf{H}_{k,l} \in \mathbb{C}^{N \times N} \) is the frequency-domain MIMO channel matrix, \( \mathbf{w} \in \mathbb{C}^{N \times 1} \) is the receiver analog combining vector, and \( n \sim \mathcal{CN}(0, \sigma_n^2) \) is additive white Gaussian noise with variance $\sigma_n^2$.

\subsection{Channel Model for ISAC}

The monostatic ISAC channel is modeled as a multipath environment, where each propagation path represents the round-trip scattering from either environmental objects or STs. We model the baseband equivalent double-directional, time-varying channel impulse response as:
\begin{equation}
\begin{split}
\label{eq:ddir}
h_l(\tau, \mathbf{\Omega}^{\text{AOA}}, \mathbf{\Omega}^{\text{AOD}}) =  h^\mathrm{BK}(\tau, \mathbf{\Omega}^{\text{AOA}}, \mathbf{\Omega}^{\text{AOD}}) + \\ \sum_{q=1}^{N_\mathrm{ST}}h_{q,l}^{\mathrm{ST}}(t,\tau, \mathbf{\Omega}^{\text{AOA}}, \mathbf{\Omega}^{\text{AOD}})
\end{split}
\end{equation}
where \( h^\mathrm{BK} \) represents the background channel component due to scattering from the static environment, and \( h_q^{\mathrm{ST}} \) models the scattering from the $q$-th ST, being $N_\mathrm{ST}$ the number of STs.

The background component \( h^\mathrm{BK} \) models the contributions from environmental scatterers that generate clutter for sensing applications. 
In this model, the environment is represented by a collection of \( N_{\mathrm{RP}} \) virtual reference points (RPs) randomly distributed around the monostatic BS, with their locations sampled from a scenario-dependent 2D distance and height distribution~\cite{10592557}. Each RP generates an independent non line-of-sight (NLoS) channel realization based on \cite{3gppTR38901}, with distinct large-scale parameters such as delay spread and angular spread. The background channel is synthesized as the sum of all RP contributions:
\begin{equation}
h^\mathrm{BK} = \frac{1}{\sqrt{N_{\mathrm{RP}}}} \sum_{r=1}^{N_{\mathrm{RP}}} h^{\mathrm{BK}}_r,
\end{equation}
where  \( h^{\mathrm{BK},r} \) is the channel contribution from the \( r \)-th RP.

The channel  \( h_q^{\mathrm{ST}} \) captures the contributions from STs, where each ST is modeled as a monostatic round-trip propagation between the BS and the \( q \)-th target. 
For each BS-ST link, a propagation condition (LoS or NLoS) is first assigned based on the height-dependent LOS probability model defined in 3GPP TR~36.777\,\cite{3gppTR36777}. Depending on the condition, the one-way path loss $ \mathit{PL}_{q,\mathrm{BS}}$ is then calculated using the corresponding scenario-specific 3GPP path loss model.
The effective round-trip path loss is $\mathit{PL}_q =2\mathit{PL}_{q,\mathrm{BS}}$.
The  magnitude associated with the \( q \)-th ST is:
\begin{equation}
a_q = 10^{ -\mathit{PL}_q/20} \sqrt\frac{4\pi\sigma_{M,q}}{\lambda^2}
\end{equation}
where \( \sigma_{M,q} \)  
is a the mean RCS value for the  the $q$-th ST  and \( \lambda \) is the carrier wavelength.
 The $q$-th ST's complex time-dependent amplitude, incorporating Doppler shift \( f_{D,q} \), is:
\begin{equation}
h^{\mathrm{ST}}_{q,l} = a_q e^{j 2\pi f_{D,q} l}
\end{equation}

The frequency-domain MIMO channel matrix \( \mathbf{H}_{k,l} \) at subcarrier \( k \) and OFDM symbol \( l \) is obtained by the Discrete Fourier transform (DFT) of the impulse response that we spatially project onto the array responses:
\begin{equation}
\begin{split}
\mathbf{H}_{k,l} =
&\sum_{r=1}^{N_\mathrm{RP}} h^{\mathrm{BK}}_r \, e^{-j 2\pi k K\Delta f \tau_r} \, \mathbf{a}_{\text{rx}}(\mathbf{\Omega}_r^{\text{AOA}})\,\mathbf{a}_{\text{tx}}^H(\mathbf{\Omega}_r^{\text{AOD}}) \\
+ &\sum_{q=1}^{N_\mathrm{ST}} h^{\mathrm{ST}}_{q,l} \, e^{-j 2\pi k K\Delta f \tau_q} \, \mathbf{a}_{\text{rx}}(\mathbf{\Omega}_q^{\text{AOA}})\,\mathbf{a}_{\text{tx}}^H(\mathbf{\Omega}_q^{\text{AOD}})
\end{split}
\end{equation}
where  \( \Delta f \) is the subcarrier spacing; \( \tau_r \) and \( \tau_q \) are the delays; \( \mathbf{a}_{\mathrm{tx}}(\cdot) \) and \( \mathbf{a}_{\mathrm{rx}}(\cdot) \) are the transmit and receive array steering vectors; and \( a_r \), \( a_{q,l} \) are the complex path gains from background and target scatterers, respectively.

\section{5G NR Radar Signal Processing Chain}

In this section, we describe the radar signal processing chain used to estimate the ST parameters, including AoA, range, and position, based on the received PRS signals.

\subsection{Angle of Arrival Estimation with Clutter Suppression}

The first step in ST localization is to estimate the AoAs of candidate STs, enabling the radar to steer the beam toward potential targets and maximize the likelihood of obtaining accurate position estimates.
Since the received signal contains contributions from both moving STs and static environmental clutter, a clutter suppression step is first applied.
We assume that the static clutter components are temporally stationary across the PRS symbols transmitted in a given analog beam direction. Under this assumption, subtracting the empirical mean across time removes the stationary background contributions while preserving the dynamic signal components reflected from the ST.
Given the frequency-domain channel matrix \( \mathbf{H}_{k,l} \) at subcarrier \( k \) and PRS symbol \( l \), the clutter-suppressed channel is computed as:
\begin{equation}
\widetilde{\mathbf{H}}_{k,l}(\theta,\phi) = \mathbf{H}_{k,l}(\theta,\phi) - \mathbb{E}_l[\mathbf{H}_{k,l}(\theta,\phi)],
\end{equation}
where $\mathbb{E}_{l}[\mathbf{H}_{k,l}(\theta,\phi)]$ denotes the empirical mean across the $l$ OFDM symbols.  
Note that \( l \) spans multiple PRS symbols transmitted with the same analog beamforming direction $ (\theta,\phi)$
providing repeated observations in time that enable clutter suppression through temporal averaging.
$\theta$ and $\phi$ respectively denote the azimuth and elevation angles for the both the AoA and the AoD. Because the BS is using monostatic sensing, the transmit path is the same as the receive path and the AoA and AoD are the same.

To identify potential STs, a peak detection algorithm can be applied over the angular domain to reveal multiple dominant arrival directions. In this paper, we focus on the detection in a single ST case. Extension to multi-target cases is deferred to future work.
Accordingly,
since the received signal power is maximized when the beam is aligned with the direction of the ST, the AoA is estimated by identifying the beamforming direction $\mathbf{w}$ that yields the highest received power:
\begin{equation}
\begin{aligned}
\bar{\mathbf{w}} = \bar{\mathbf{f}} = \argmax_{\mathbf{w,f} \in \mathcal{C}} \sum_{k,l} \left| \mathbf{w}^H \widetilde{\mathbf{H}}_{k,l} \mathbf{f} \right|^2 
\end{aligned}
\end{equation}
where \( \mathcal{C} \) denotes the analog beamforming codebook.
This optimization is performed through a beam-sweeping procedure compatible with 5G NR beam management operations, such as those defined for PRS-based positioning, where PRS resources are transmitted over multiple beams to facilitate angle estimation \cite{3gppTR38855}. 
The resulting beamforming vectors \( (\bar{\mathbf{w}}, \bar{\mathbf{f}}) \) identify the estimated angle of arrival \( \widehat{\mathbf{\Omega}} = (\widehat{\theta}, \widehat{\phi}) \), where \( \widehat{\theta} \) and \( \widehat{\phi} \) denote the estimated azimuth and elevation angles, respectively.

\subsection{Effective Channel Estimation}

Following beam alignment toward the ST using analog beamforming, the effective channel estimation is performed at each active subcarrier.
To extract the effective scalar channel \( \widehat{h}_{k,l} \), matched filtering with the known conjugate PRS symbol $s_{k,l}^*$ is applied:
\begin{equation}
\label{eq:channel_estimation}
\widehat{h}_{k,l} = s_{k,l}^* y_{k,l}  \approx h^{\mathrm{ST}}_{l} \, e^{-j 2\pi k K\Delta f \tau}+n.
\end{equation}
Here, the index \( q \) is dropped from \( h^{\mathrm{ST}}_{q,l}\) and \( \tau_q \) since we assume a single  ST is present in the scene. The approximation accounts for residual clutter interference and imperfect analog beamforming, which may cause leakage from weaker paths.

\subsection{Range Estimation}

The estimated \( \widehat{h}_{k,l} \) serves as the input to the range estimation procedure.  
Range estimation is performed by exploiting the phase shifts across PRS subcarriers due to propagation delays to and from the STs.
To estimate the delay of the ST, a range-processing  IDFT of length $N_R$ is applied:
\begin{equation}
\label{eq:dft}
r_l(d) = \frac{1}{N_R} \sum_{k=0}^{N_R-1} \widehat{h}_{k,l} \mathrm{e}^{ j 2\pi \frac{k d}{N_R} },
\end{equation}
where 
\( d \) is the delay bin index.
Substituting eq. (\ref{eq:channel_estimation}) into eq. (\ref{eq:dft}):
\begin{equation}
r_l(d) = h^{\mathrm{ST}}_{l} \left( \frac{1}{N_R} \sum_{k=0}^{N_R-1} \mathrm{e}^{\left( j 2\pi k \left( \frac{d}{N_R} - K \Delta f \tau \right) \right)} \right) + n.
\end{equation}
The IDFT sum forms a discrete Fourier series that achieves a peak when
$d_{\text{peak}}/N_R = K \Delta f \tau$.
Thus, the round-trip delay is estimated as:
$\hat{\tau} =  d_{\text{peak}}/(N_R K \Delta f)$,
and the corresponding estimated range \( R \) is:
\begin{equation}
\hat{R} = \frac{c \hat{\tau}}{2} = \frac{c}{2} \frac{d_{\text{peak}}}{N_R K \Delta f},
\end{equation}
where \( c \) is the speed of light in the medium.

\subsection{Detection Decision}
Although the delay bin $d_{\text{peak}}$  corresponding to the maximum of the range profile has been identified, it is necessary to verify whether this peak corresponds to an actual ST or is the result of a false alarm due to noise or residual clutter.
The detection statistic is defined as the peak-to-average ratio (PAR) across the range bins:
\begin{equation}
\text{PAR} = \frac{ |r_l(d_{\text{peak}})|}{\frac{1}{N_R} \sum_{d} |r_l(d)|}.
\end{equation}
If the PAR exceeds a predefined threshold \( \eta \), a detection is declared. Otherwise, the observation is considered as noise or clutter.  
The choice of the detection threshold \( \eta \) controls the trade-off between detection probability and false alarm probability and can be tuned according to the desired operating point.
If detection is declared, \( d_{\text{peak}} \)  is used to estimate the ST range and 3D position.

\subsection{3D Target Position Estimate}

Given the estimated range \( \widehat{R} \) and angle of arrival \( \widehat{\mathbf{\Omega}}  = (\widehat{\theta} , \widehat{\phi} ) \), the unit direction vector is defined as \( \mathbf{u} = \left( \cos(\widehat{\theta} )\cos(\widehat{\phi} ), \, \sin(\widehat{\theta} )\cos(\widehat{\phi} ), \, \sin(\widehat{\phi} ) \right)^T \).
The estimated ST position \( \widehat{\mathbf{x}} \in \mathbb{R}^3 \) is then given by
$\widehat{\mathbf{p}}_\mathrm{ST} = \mathbf{p}_{\text{BS}} + \widehat{R} \, \mathbf{u},$
where $\boldsymbol{p}_{\mathrm{BS}}$ 
denotes the known location of the BS.

\section{5G NR Radar Performance Evaluation}
This section presents a performance evaluation of the proposed 5G\,NR radar system for airborne ST sensing. We first describe the simulation setup, including the parameters and propagation models adopted for different deployment scenarios. We then present the detection and position error results for different UAV altitudes on the system’s sensing performance.

\subsection{Scenarios and Simulation Setup}
\begin{table}[t]
\centering
\caption{Deployment parameters for UMi-AV and UMa-AV scenarios based on 3GPP TR 36.777.}
\label{tab:scenario_parameters}
\begin{tabular}{l@{\hskip 2pt}c@{\hskip 2pt}c@{\hskip 2pt}c}
\toprule
\textbf{Parameter} & \textbf{UMi-AV} & \textbf{UMa-AV} \\
\midrule
BS height (m) & 10 & 25  \\
ISD (m) & 200 & 500 \\
LOS probability & - & + \\
$\sigma_{SF,\text{LoS}}$ (dB) & $\max(5e^{-0.01 h_{\text{UAV}}},2)$ & $4.64e^{-0.0066 h_{\text{UAV}}}$  \\
$\sigma_{SF,\text{NLoS}}$ (dB) & 8 & 6  \\
\bottomrule
\end{tabular}
\end{table}
We consider two standard 3GPP deployment environments, UMi-AV and UMa-AV, as defined in TR~36.777~\cite{3gppTR36777}. The key deployment parameters are summarized in Table~\ref{tab:scenario_parameters}. The UMi-AV scenario models dense urban areas with small cells, featuring a BS height of 10\,m and a dense inter-site distance (ISD) of 200\,m. UMa-AV captures larger urban deployments with macrocell BSs at 25\,m height and ISD of 500\,m. 
The two scenarios adopt different LoS probability models. LoS refers to the existence of an unobstructed direct path between the BS and the UAV ST. The LoS probability follows the aerial useq equipment (UE) height-dependent modeling approach introduced in Table~B-1 in 3GPP TR~36.777~\cite{3gppTR36777}. In UMa-AV, since the BS antennas are positioned above rooftop levels, a 100\,\% LoS probability is assumed above UAV height $h_\mathrm{UAV} =100$\,m. In contrast, in UMi-AV, where the BS antennas are located below rooftop height, NLoS conditions are more prevalent, and no upper height threshold for guaranteed LoS is defined.
Additionally, each scenario applies different shadow fading (SF) standard deviations for LoS and NLoS conditions. The SF standard deviation is larger in the UMi-AV scenario than in the UMa-AV scenario. 
This reflects the increased signal variability typical of dense urban environments, where a rich scattering environment, densely packed, tall buildings, and abundant obstructions result in more pronounced shadowing effects.
In contrast, macrocell deployments experience less variability due to the higher BS placement and the more open surroundings. 

The simulation assumes a carrier frequency of \( f_c = 30 \,\mathrm{GHz} \) and a PRS allocated into $N_\mathrm{Grid} = 66$ resource blocks (RBs) with numerology \( \mu = 3 \), i.e., subcarrier spacing \( \Delta f = 120 \,\mathrm{kHz} \). This configuration results in an overall bandwidth of 100\,MHz.
PRS symbols are transmitted using a comb pattern with \( K = 4 \), and each PRS occasion consists of \( L_\mathrm{PRS} = 4 \) OFDM PRS symbols. A total of \( N_\mathrm{PRS} = 256 \) PRS occasions are simulated.
The BS is equipped with a $32\times32$ uniform planar array. The total transmit power is adjusted to comply with FCC regulations, ensuring a maximum equivalent isotropically radiated power (EIRP) of 75~dBm per 100~MHz bandwidth.

 For each deployment scenario, one ST representing a UAV is uniformly dropped across multiple independent realizations within a single cell, generating a total of 4000 ST drops per configuration. We consider four ST heights, specifically $h_\mathrm{UAV}\in\{25\,\mathrm{m}, 50\,\mathrm{m}, 100\,\mathrm{m},\  200\,\mathrm{m}\}$, to capture the  detection performance variation with altitude. Once the ST position is randomly deployed, the corresponding link conditions are generated: the LoS or NLoS condition is determined probabilistically, followed by the computation of the path loss and application of SF. The ST RCS is set to $\sigma_M = -12.81$~Decibels relative to a square meter (dBsm), reflecting values proposed in ongoing 3GPP standardization efforts for small UAVs. The background channel has been realized by dropping $N_\mathrm{RP} = 3$ RPs.

\subsection{Detection and Position Error Performance}

\begin{figure}
\vspace{3mm}
    \centering
    \subfloat[False alarm probability vs $\eta$.]{
        \includegraphics[width=0.45\columnwidth]{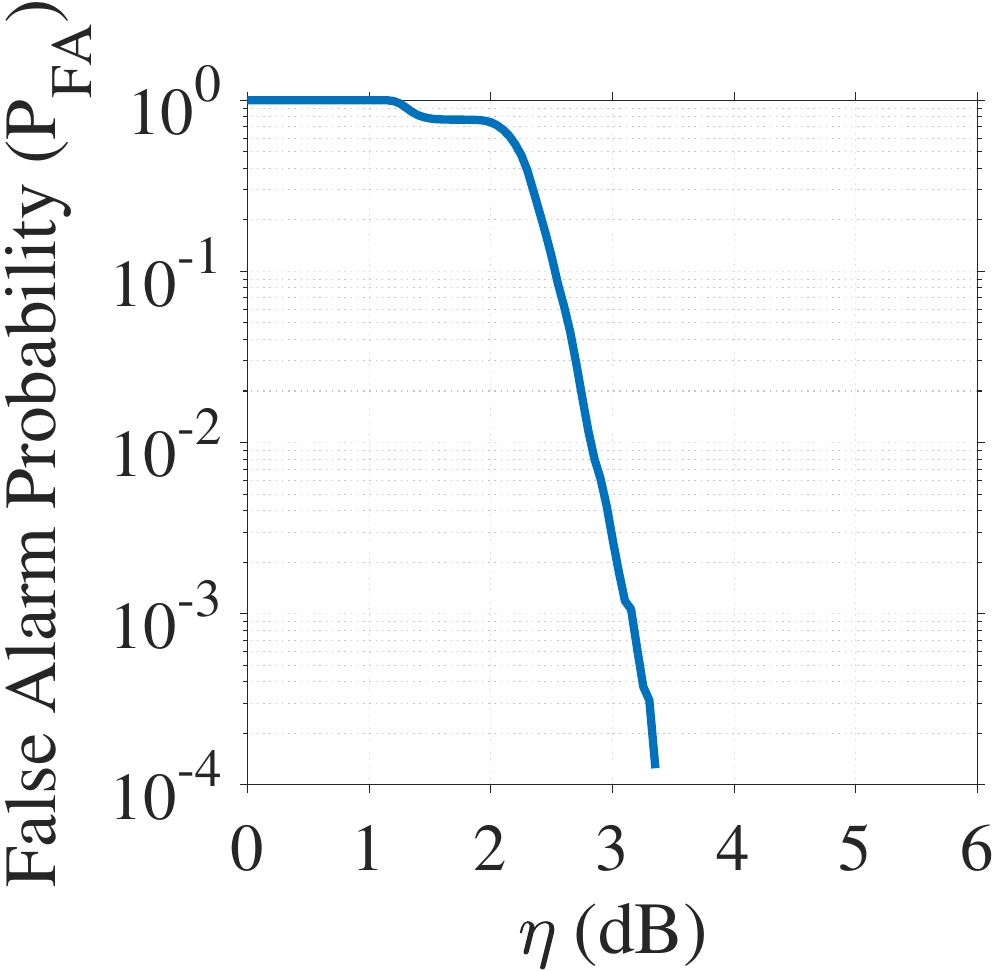}
        \label{fig:roc_pfa}
    }
    \subfloat[Detection probability vs  $\eta$.]{
        \includegraphics[width=0.45\columnwidth]{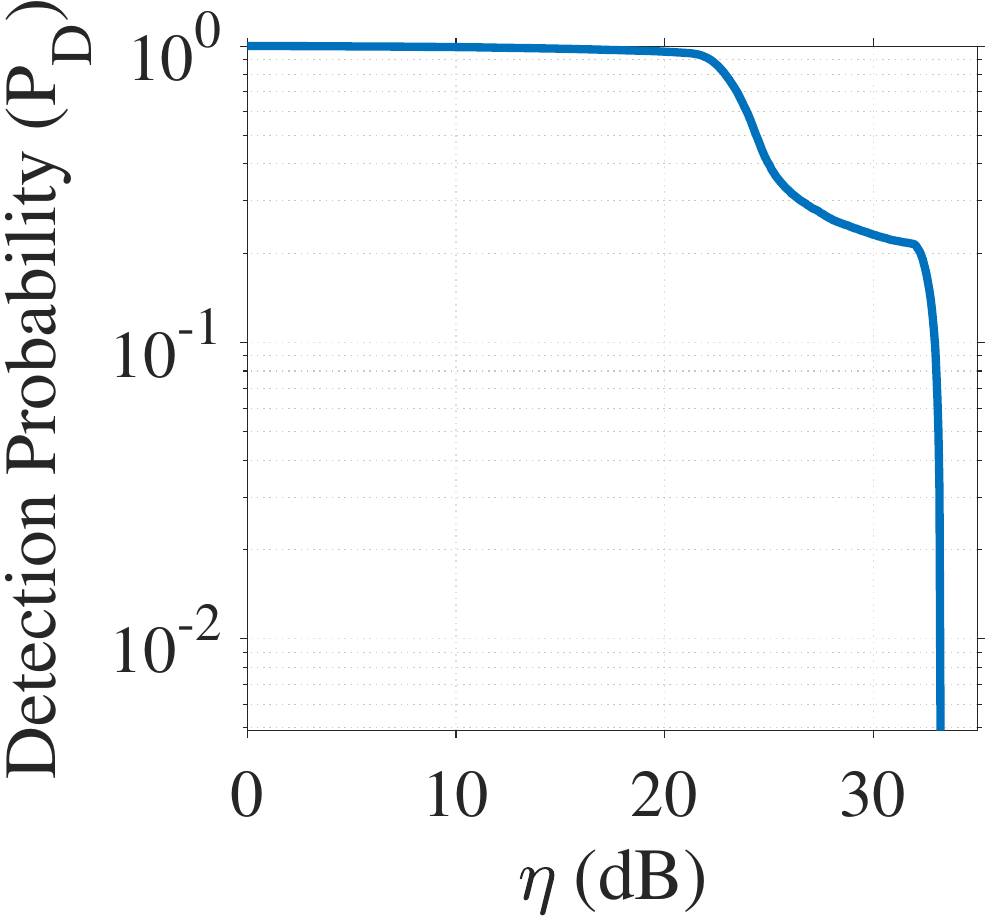}
        \label{fig:roc_pd}
    }
    \caption{5G NR Radar operating curves.}
    \label{fig:performance_metrics}
\end{figure}
To determine a suitable detection PAR threshold 
$\eta$, we simulate the detection behavior across all considered configurations. Specifically, we run simulations for each of the 8 combinations of deployment scenario (UMi-AV, UMa-AV) and UAV altitude (25\,m, 50\,m, 100\,m, 200\,m), both in the presence and absence of a target, resulting in a total of 64\,000 samples. The false alarm probability and detection probability are shown as functions of $\eta$ in Fig. \ref{fig:roc_pfa} and Fig. \ref{fig:roc_pd},  respectively. A detection threshold of $\eta = 3.4$\,dB provides a good trade-off between high detection rates and low false alarm probabilities across all tested configurations and is adopted for the subsequent evaluation.

\begin{figure*}
    \centering
    \subfloat[UMi-AV, $h_\mathrm{UAV}= 25\,\mathrm{m}$]{
        \includegraphics[width=0.22\textwidth]{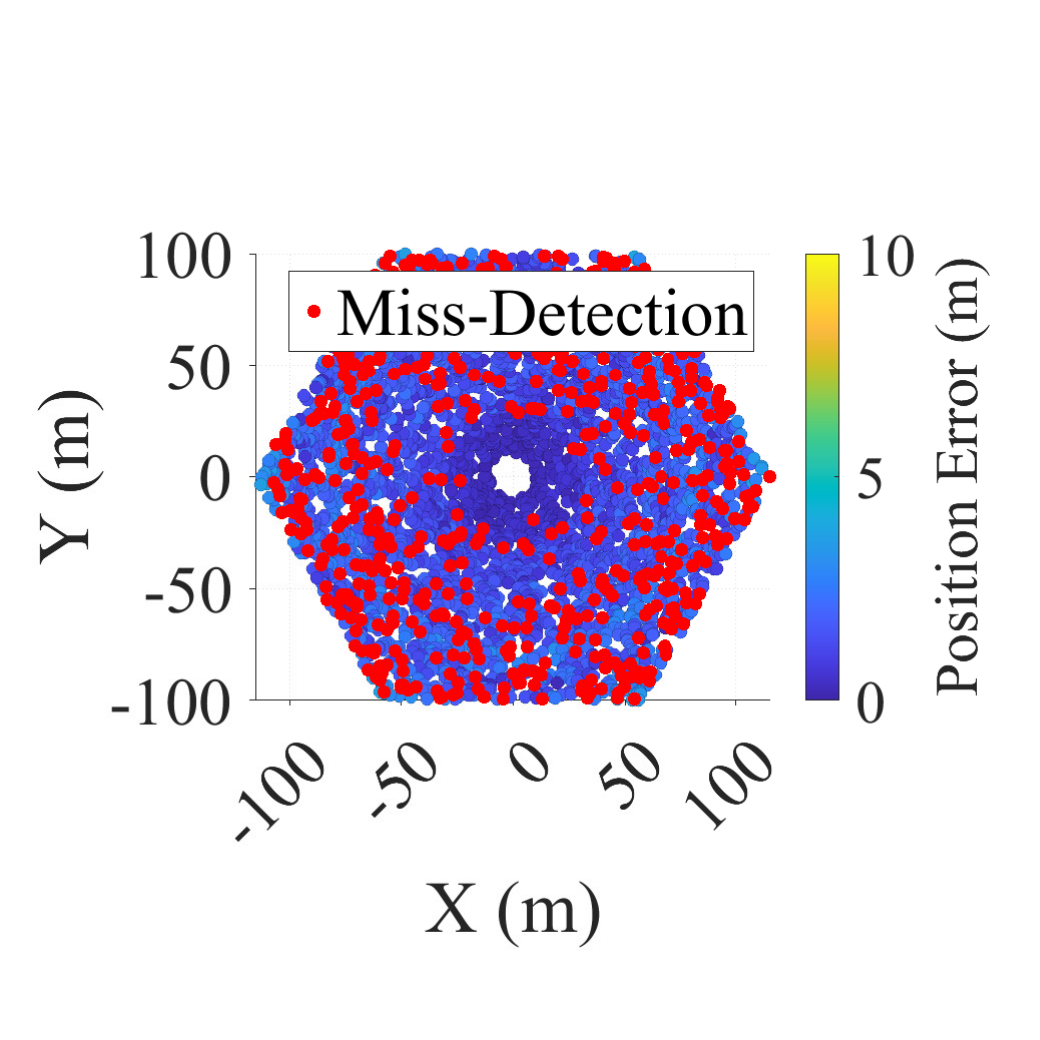}
        \label{fig:UMi25}
    }
    \hspace{1mm}
    \subfloat[UMi-AV, $h_\mathrm{UAV}= 50\,\mathrm{m}$]{
        \includegraphics[width=0.22\textwidth]{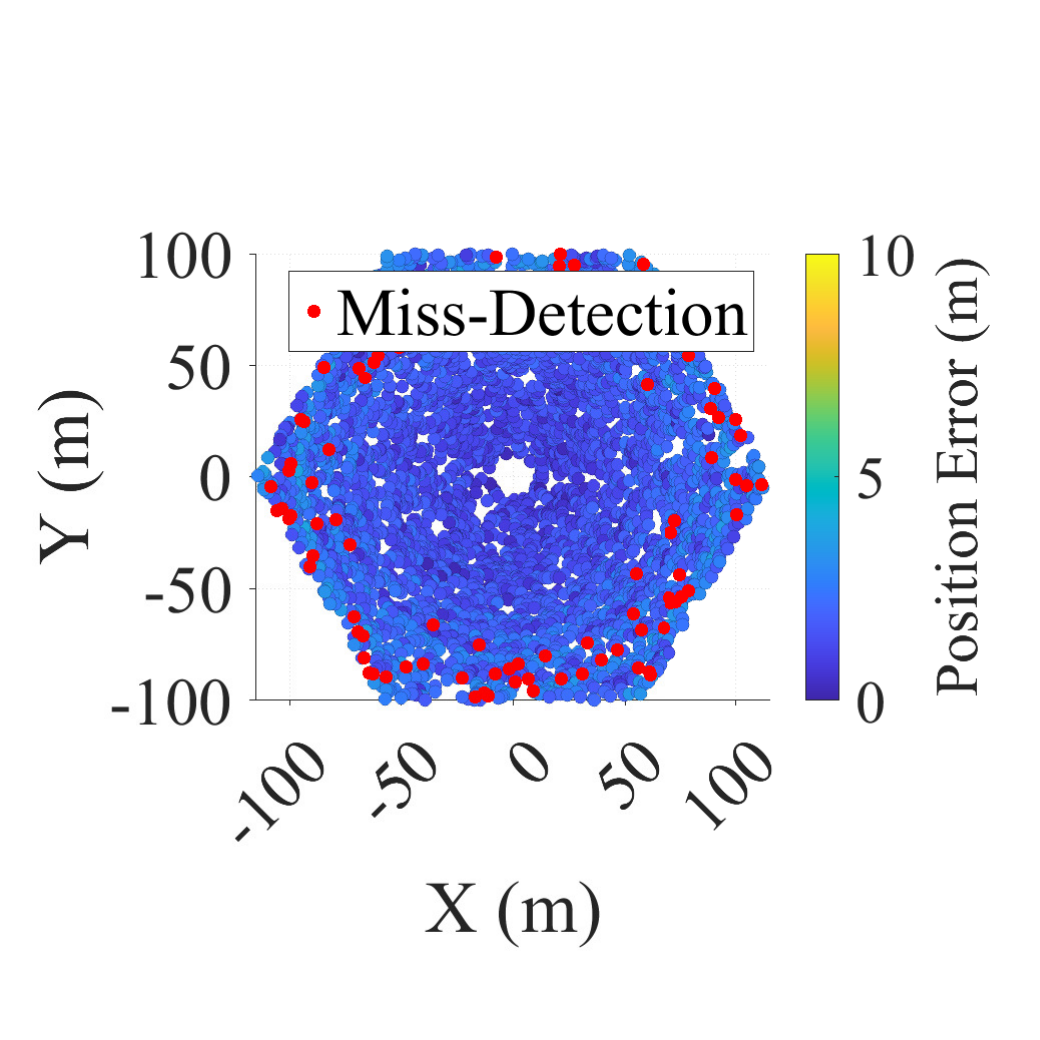}
        \label{fig:UMi50}
    }
    \hspace{1mm}
    \subfloat[UMi-AV, $h_\mathrm{UAV}= 100\,\mathrm{m}$]{
        \includegraphics[width=0.22\textwidth]{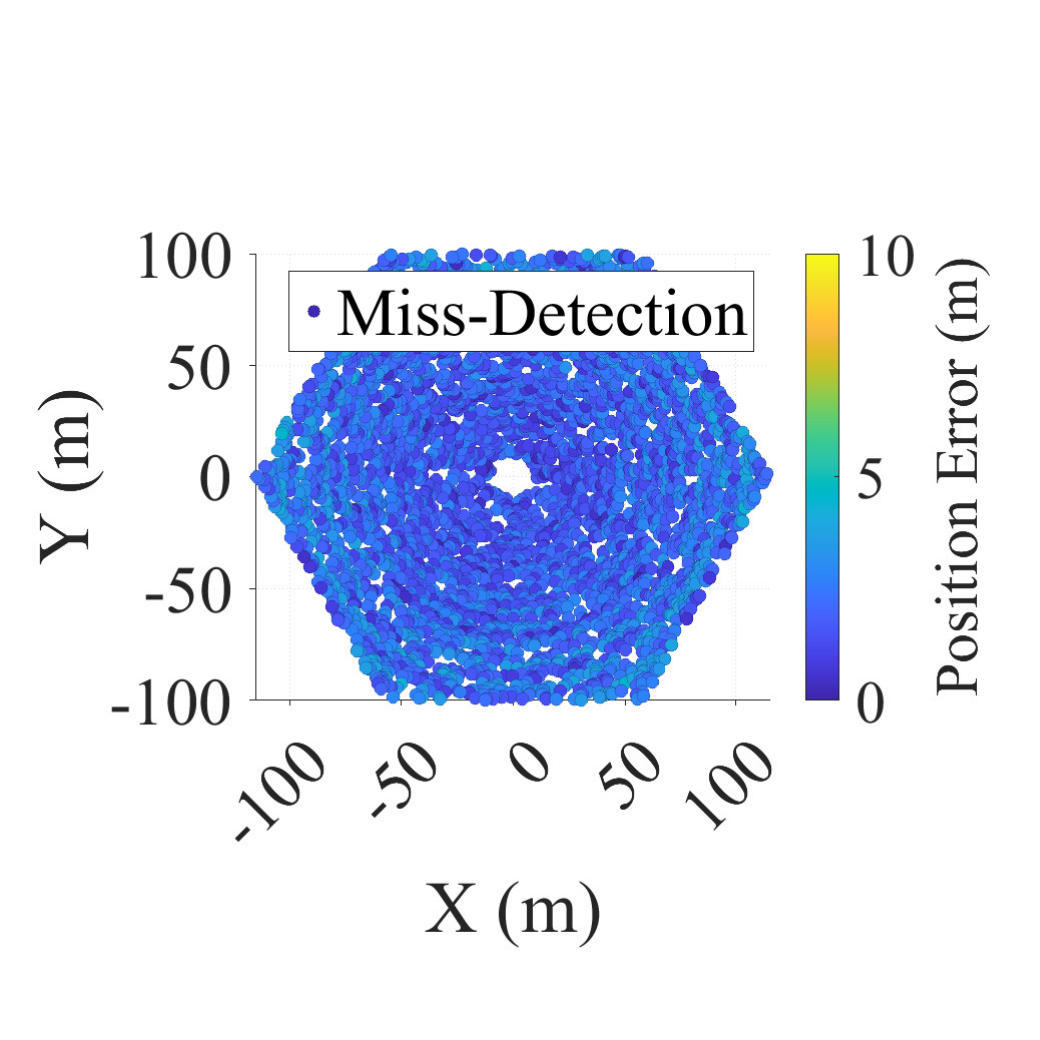}
        \label{fig:UMi100}
    }
    \hspace{1mm}
    \subfloat[UMi-AV, $h_\mathrm{UAV}= 200\,\mathrm{m}$]{
        \includegraphics[width=0.22\textwidth]{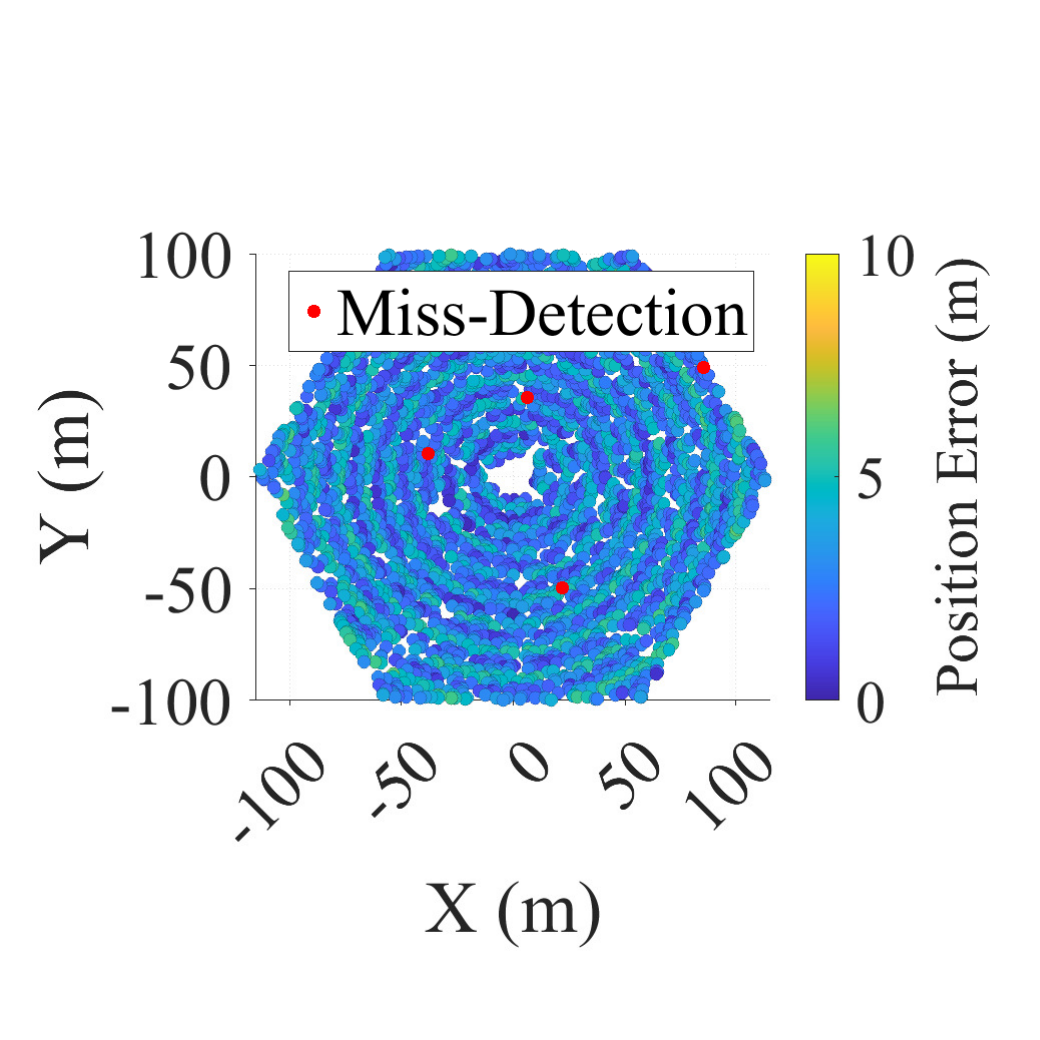}
        \label{fig:UMi200}
    }
    \\
    \subfloat[UMa-AV, $h_\mathrm{UAV}= 25\,\mathrm{m}$]{
        \includegraphics[width=0.22\textwidth]{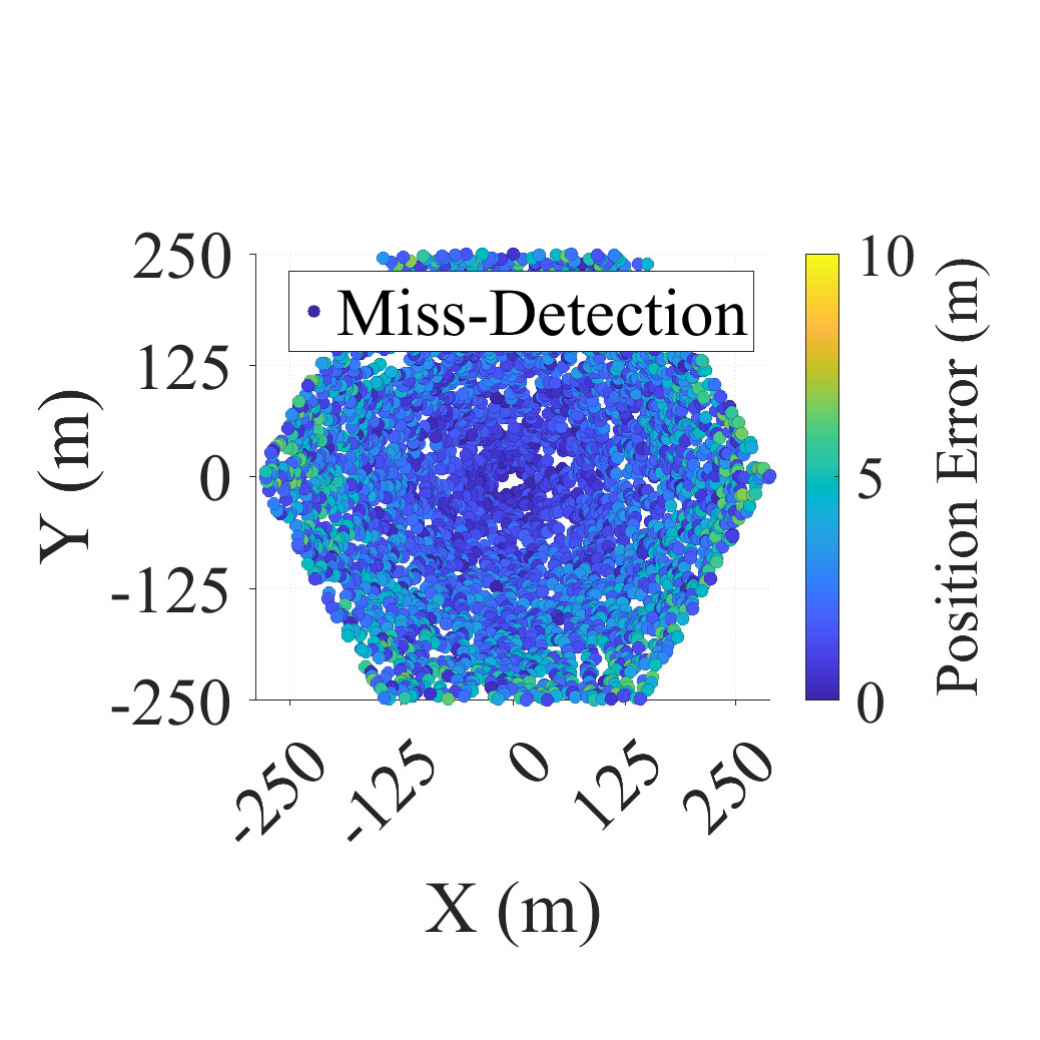}
        \label{fig:UMa25}
    }
    \hspace{1mm}
    \subfloat[UMa-AV, $h_\mathrm{UAV}= 50\,\mathrm{m}$]{
        \includegraphics[width=0.22\textwidth]{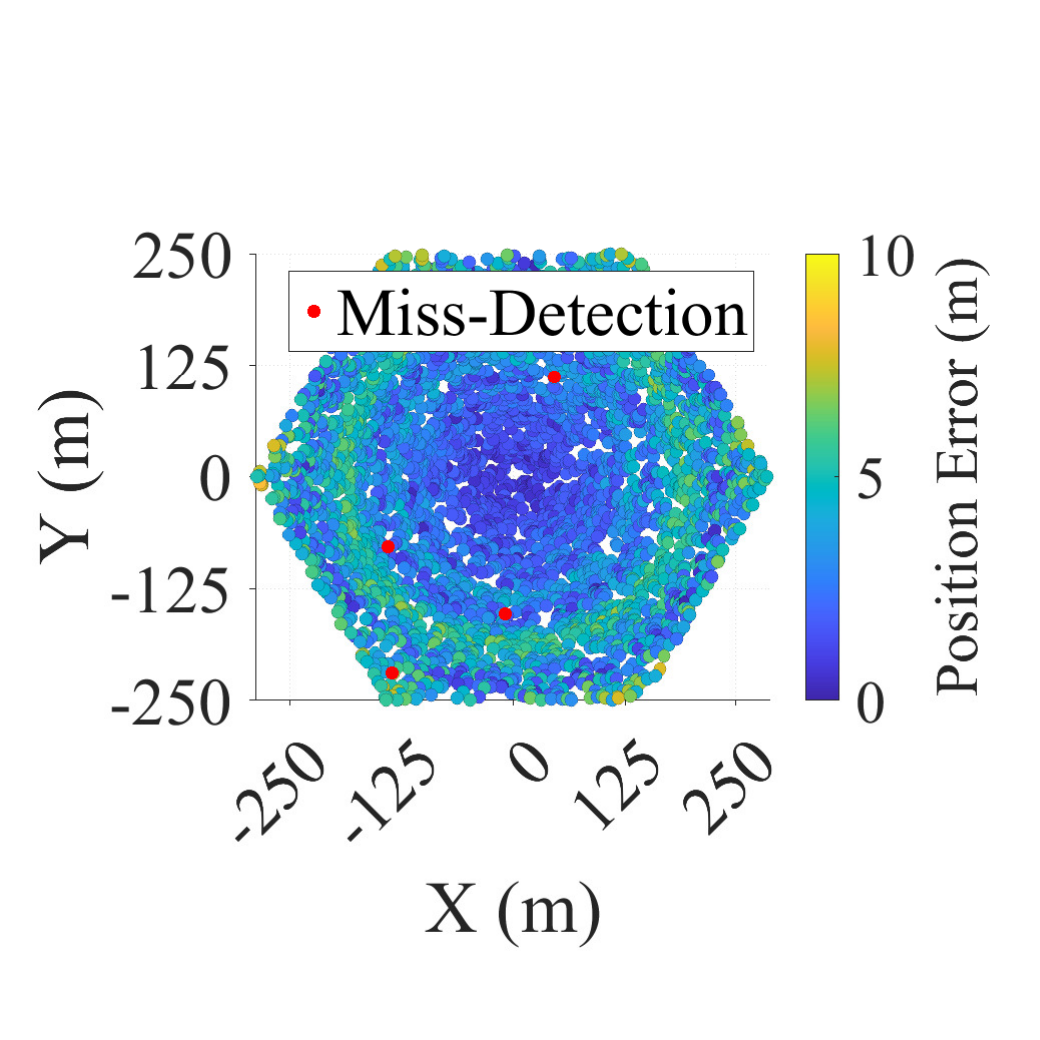}
        \label{fig:UMa50}
    }
    \hspace{1mm}
    \subfloat[UMa-AV, $h_\mathrm{UAV}= 100\,\mathrm{m}$]{
        \includegraphics[width=0.22\textwidth]{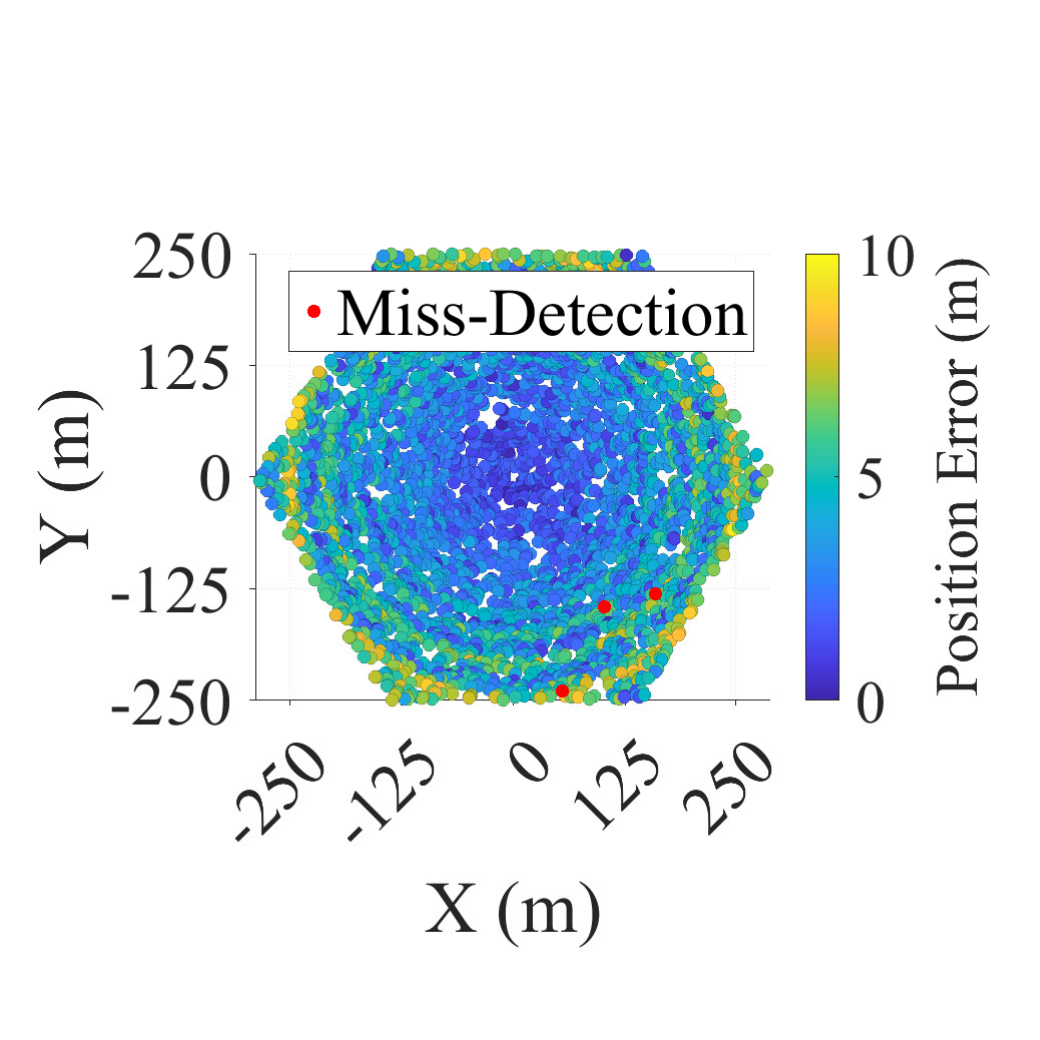}
        \label{fig:UMa100}
    }
    \hspace{1mm}
    \subfloat[UMa-AV, $h_\mathrm{UAV}= 200\,\mathrm{m}$]{
        \includegraphics[width=0.22\textwidth]{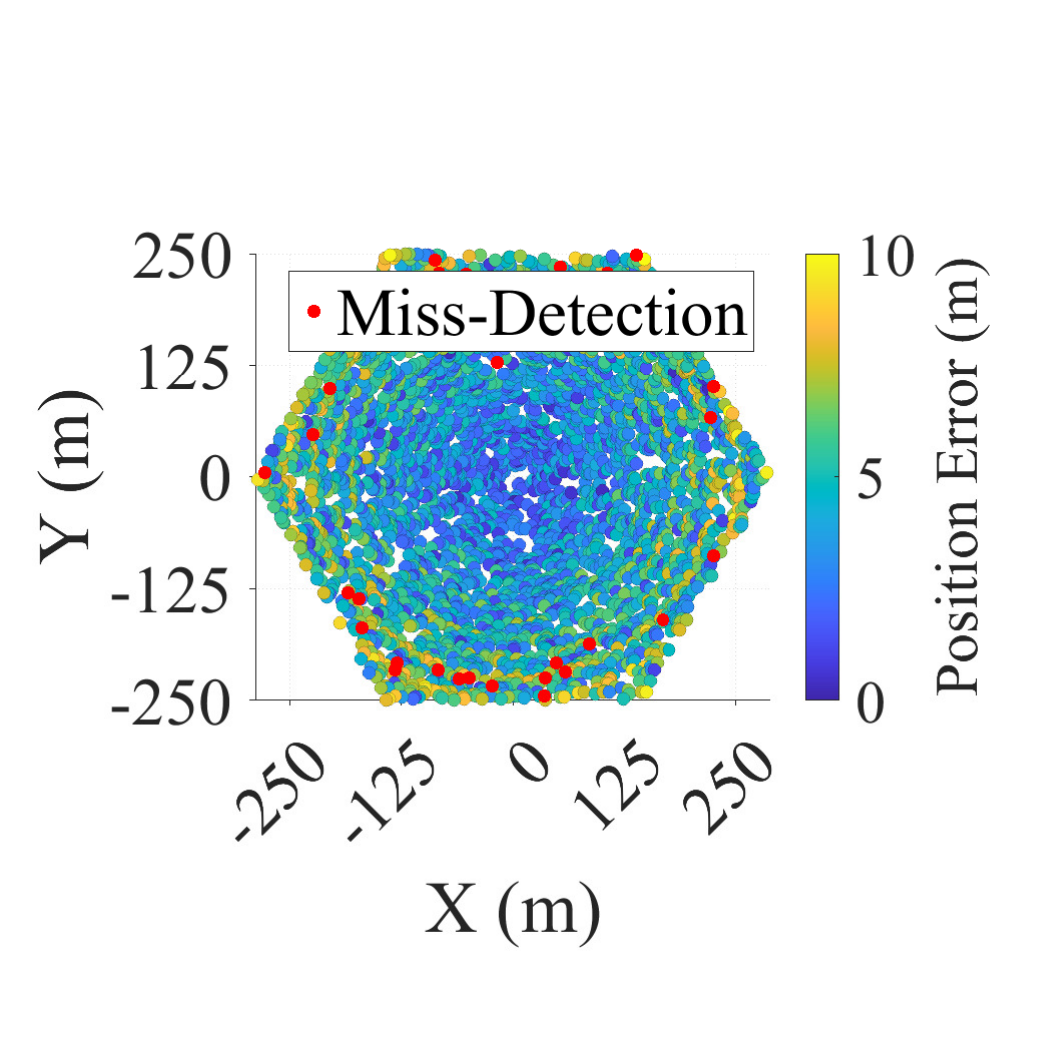}
        \label{fig:UMa200}
    }    
    \caption{Detection and positioning performance across two deployment scenarios (UMi, UMa) at different UAV altitudes (25 m, 50 m, 100 m, 200 m).}
    \label{fig:allResults}
\end{figure*}

The position estimation error is evaluated for each UAV drop using the radar processing chain detailed in Section~IV. Detection and position error performance across different deployment environments and UAV heights are shown in Fig.~\ref{fig:allResults}. Each subplot corresponds to a specific UAV altitude for the considered UMi-Av and UMa-Av scenarios.
 The UMi-AV scenario exhibits significantly higher miss detection rates compared to UMa-AV, primarily due to dense urban clutter. The most challenging case is observed at 25\,m altitude, where the ST is low and more likely to be blocked. At 50\,m, detection performance improves, with fewer missed detections, and becomes reliable at 100\,m and 200\,m, where the UAV is sufficiently elevated above the dominant clutter layer, resulting in almost no missed detections.
In contrast, the UMa-AV scenario demonstrates very high detection performance at low altitudes, particularly at 25\,m, where no missed detections are observed. This is attributed to the elevated BS height and the open macrocell environment. However, as UAV altitude increases, some missed detections begin to appear, especially at 100\,m and 200\,m. At larger ST distances, lower SNR leads to occasional detection failures.
Moreover, we notice that position estimation degrades with increasing UAV altitude. This is due to the growing distance between the BS and the ST.
Although the array's angular resolution is fixed, the cross-range uncertainty increases with distance, as small angular deviations translate into larger absolute position errors at greater ranges.
When comparing the two deployment types, UMi-AV achieves lower position error than UMa-AV at corresponding UAV heights. This is a direct consequence of the denser deployment and shorter BS–ST distances in UMi-AV.

\begin{figure}
    \centering
    \subfloat[Miss detection probability across scenarios and UAV heights.]{
        \includegraphics[width=0.45\columnwidth]{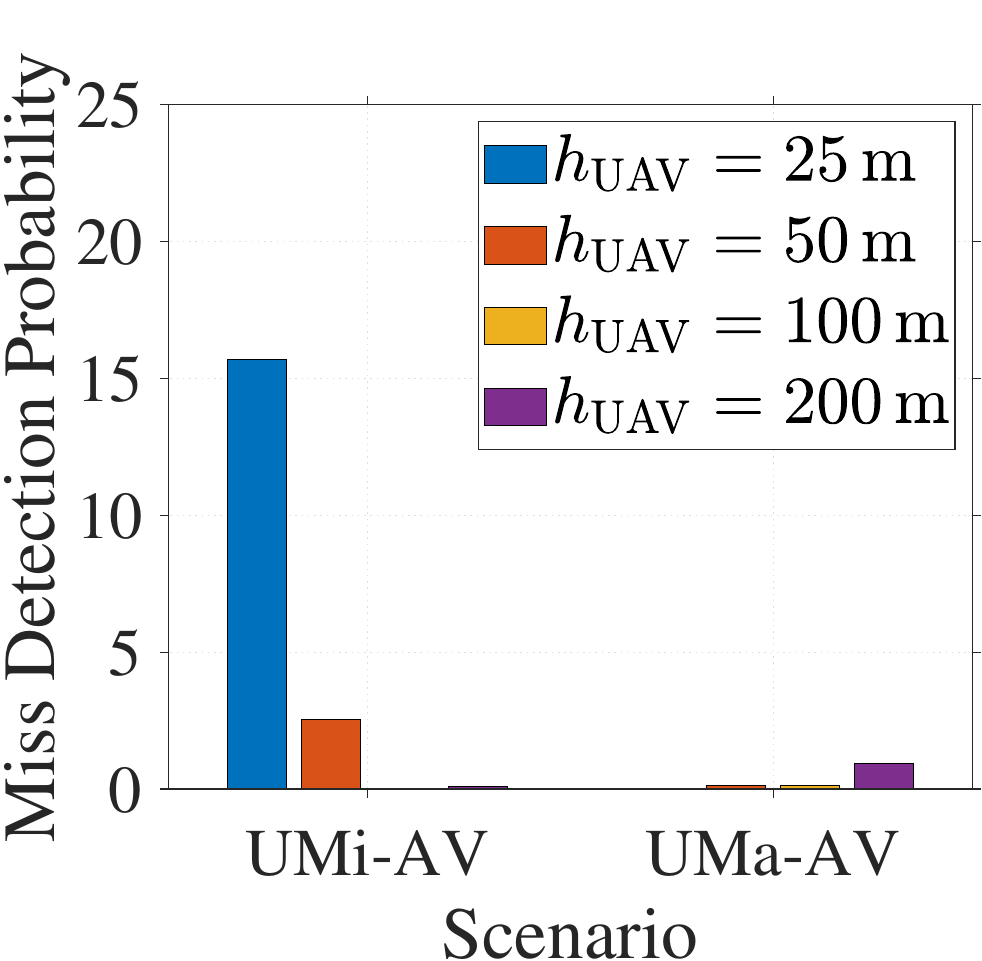}
        \label{fig:miss_detection}
    }
    \subfloat[Empirical CDF of the position estimation error.]{
        \includegraphics[width=0.45\columnwidth]{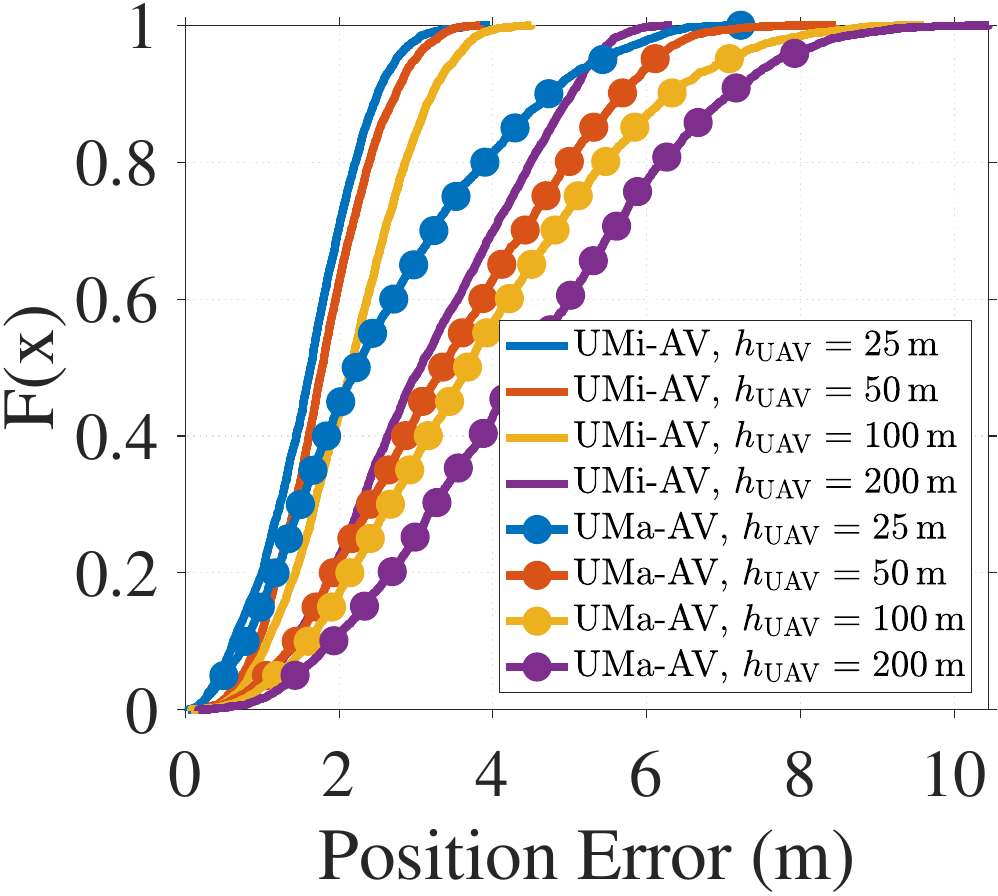}
        \label{fig:position_error_cdf}
    }
    \caption{Statistics of detection and position error in UMi-AV and UMa-AV scenarios.}
    \label{fig:stats}
\end{figure}

Statistics of the above results are presented in Fig.~\ref{fig:stats}.  
In Fig.~\ref{fig:miss_detection}, the miss detection probability is shown as a function of UAV altitude for both deployment scenarios. In the UMi-AV scenario, miss detection is particularly pronounced at low altitudes due to strong urban clutter and frequent NLoS conditions. The probability reaches approximately 16\,\% at 25\,m, decreases to around 3\,\% at 50\,m, and becomes negligible at 100\,m and 200\,m. 
In contrast, the UMa-AV scenario exhibits negligible miss detection rates below 100\,m, owing to the elevated BS placement and more open environment. However, a slight increase in miss detection appears at 200\,m altitude, reaching approximately 1\,\%, which is attributed to the reduced SNR at greater BS–ST distances.


Figure~\ref{fig:position_error_cdf} presents the empirical cumulative distribution function (CDF) of the position estimation error, conditioned on successful detections, across different UAV altitudes and deployment scenarios. In both scenarios, position error degrades with increasing UAV altitude due to the growing BS–ST distance.
In UMi-AV, the best performance is achieved at 25\,m, with errors within 2\,m in most cases. As altitude increases to 50\,m, 100\,m, and 200\,m, the error distributions progressively degrade, with errors within 4\,m. A similar trend appears in UMa-AV, with increasing position errors at higher altitudes, with position errors within 8\,m.
Across all altitudes, position estimation in UMi-AV is better than in UMa-AV due to its denser deployment and shorter BS–ST distances. 


\section{Conclusion}
This paper demonstrates the feasibility of using standardized 5G\,NR PRS for UAV detection and localization in realistic urban propagation environments. A complete radar processing chain is implemented, including clutter suppression, angle and range estimation, and 3D position reconstruction. Simulation results across UMi-AV and UMa-AV scenarios show that detection and positioning performance are highly sensitive to deployment geometry and UAV altitude. Dense UMi deployments present greater clutter-induced challenges, especially at low UAV altitudes, but offer higher position accuracy due to shorter BS–ST distances. Conversely, UMa deployments achieve more robust detection at low altitudes but exhibit larger position errors at larger UAV distances.
Future work focuses on extending the analysis to multi-target detection and tracking, and on exploiting multiple cooperating BSSs to improve robustness and mitigate NLoS limitations in urban environments.


\end{document}